\documentclass[12pt]{article}
\usepackage{graphicx}

\begin{document}

\title{On the ontological ambiguity of Physics facing Reality}

\author{J.E. Horvath$^{*{1}}$, R. Rosas Fernandes$^{1}$ and T.E. Idiart$^{1}$\\
{Departamento de Astronomia}, {IAG-USP}, {S\~ao Paulo SP}, {Brazil}}

\maketitle

\abstract{We discuss in this work the contradictory position in modern Physics between the existence of a microphysical quantum Reality and macrophysical classical one. After discussing some basic concepts in Philosophy, we revisit the situation of Quantum Mechanics results, particularly after the confirmation of violations of Bell's Theorem, and its significance to create a tension between the micro and the macro world, which is very fundamental and unsolved.}

\section{Introduction}\label{sec1}

\subsection{Atomism and the natural world}

When the pre-Socratic philosopher Leucippus ($\sim$ 450 a.C.) maintained that matter is constituted of elementary entities called atoms, he did so within a philosophically-oriented attitude and without any hard evidence which we would call nowadays ``scientific''. We do not know much about Leucippus, he would have written two books entitled {\it The Great World System} and {\it On Mind}. Besides the introduction of the concept of atom, as indivisible, partless entities that gather, entangle and disperse within a vacuum, in a whirl, Leucippus expresses that

\bigskip
{\it Nothing happens at random, but everything from reason and by necessity}

\bigskip
Leucippus, {\it On Mind}

\bigskip
The lack of further information does not allow us to elaborate further his ideas, although they survived in a mixed and more developed form in the texts by his disciple, Democritus of Abdera. It is important to stress again that in Ancient Greece times there was no evidence that the atoms existed [4]. In its beginning, we can consider the Atomist doctrine as a Metaphysical thought. Two millennia had to pass to prove experimentally the atomic hypothesis, even in a modified form reprocessed a few times along History.

As a philosophical doctrine, the idea that atoms constitute all existing matter was strongly rejected by Plato, Aristotle, neoplatonists, paripathetics, all Christian philosophers and also by rationalistic thinkers, Descartes among many others. It is difficult to find a Philosophy course in which a great deal of attention is paid to atomistic philosophers, in spite of modern evidence for the discrete nature of matter. The introduction to pre-Socratic thinking is usually focused initially on Parmenides and his book On the Order of Nature, and in particular on his statement ``being is and not-being is not'', actually one of the first ontological statements known. We find here the beginning of the fundamental Principle of Identity and all Western Metaphysics, reverberating all the way down to the present days.


\subsection{Platonic-Aristotelic Metaphysics}

Even if Parmenides statements can sound strange at first sight, they became stronger when Plato incorporated them, together with some Pythagorean thoughts, to his own Universe. As an example, it is undeniable that one simple mathematical addition $3 + 5 = 8$ is rational, timeless, unchangeable, perfect and therefore, according to Plato, it has more reality than any experience obtained through the senses, which can fool us [6].
Another example is a square, a geometrical figure of four equal sides and internal angles. Each time we analyze rationally a square figure, we arrive at a rational, universal and unquestionable reality. A square has always been, and will be, a timeless, unchanged, perfect figure.

Plato goes even further than Parmenides and holds in his famous Metaphysics that the phenomenological world, as  perceived through our senses, is particular, timely and changeable, imperfect and therefore unreal. As is well-known, he claims that the Ideal world, accessible by reason, is real while  the Phenomena world is not, it just contains imperfect versions, copies, of the perfect forms.
Aristotle maintains that we can access the metaphysical world of Ideas because we have experiences in the Phenomena world. Both philosophers embrace the concept of a metaphysical world with the same dogmatic posture. We stress again that the idea of perfection is essential to guarantee the reality of the metaphysical world.

Aristotle goes beyond Plato formulating the Principle of Identity, which states that a) what it is, is ; b) something can not ``be'' and ``not be'' at the same time ; and c) if something is, then it is equal to itself. The Syllogistic Logic started by Aristotle in his work {\it {\'Organon}}, is meant as a tool to reach necessary and irrefutable conclusions.

If we compare Leucippus and Democritus philosophy with the Plato and Aristotle, the most important representatives of Attic schools, it is easy to characterize both of them as {\it latu sensu} Idealistic philosophers. However, the former atomists are more difficult to classify, in spite that the few authentic fragments that survived show strong hints for an extreme Materialism. The latter statement is arguable, but to classify them as {\it physicalists} (there is nothing which is not contained in the natural world) is more certain. For example, according to Diogenes Laertius, Democritus wrote:

\bigskip
{\it By convention sweet is sweet, bitter is bitter, hot is hot, cold is cold, color is color; but in truth there are only atoms and the void}

\bigskip
{\it The souls and the mind are the same thing, and composed of a special kind of atoms}

\bigskip
Democritus of Abdera, circa 400 a.C.

\bigskip
It is worth pointing out that Leucippus-Democritus philosophy is taught quite differently in Physics and Astronomy undergraduate courses, where the scientific proof for the existence of atoms is presented, something that did not happened until the 19th century. That is, the focus of the modern teaching comprises mainly the transition from a Metaphysical atomism to the scientific one.
We now turn our attention to a brief exposition to the two categories essential for the present article: the modern Rationalist and Empiricist thoughts.

\section{The Modern Thinking}

The modern thinking [6] has many pioneers. There are arguments to consider Hugo Grotius (1583 – 1645) as the founder of modern thinking, although Francis Bacon (1561 – 1626) and Ren\'e Descartes (1596 – 1650) are also suggested. All of them can be bundled as the originators of Modern Philosophical  thinking. The Rationalism of  Ren\'e Descartes, later contrasted with the radical Empiricism of Francis Bacon (1561-1626), George Berkeley (1685 – 1753) and David Hume (1711 – 1776) are of particular interest.

\subsection{The Rationalism of René Descartes}

Besides being considered the founder of Modern Philosophy. Descartes is also considered the father of Modern Mathematics, and a key figure for the Scientific Revolution, especially in its theoretical foundations.
In 1637 Ren\'e Descartes published a work on Optics, entitled {\it La Dioptrique} (Dioptrics), but it is its preface, known as {\it Discours de la M\'ethode} (Discourse on the Method) that brought fame to his author. Later in 1641, Descartes wrote the {\it Meditationes de Prima Philosophia} (Meditations on First Philosophy) on Metaphysics in which he deepens the method of doubt and expresses his famous {\it cogito ergo sum} (I think, therefore, I exist)). The main intention of Descartes was to demonstrate rationally and metaphysically the existence of God, and he holds that the latter is an innate idea: men are born with the idea of God, independently of any previous experience. Cartesianism spreads all over Europe. Descartes, however, did not praise Galileo's observational discoveries (he stated that did not complete the reading of  {\it Siderius Nuncius}), and also rejects atomism.
With Descartes the modern notion of Subject was born, as a starting point of all the philosophical process of  searching the truth. Before Descartes, ``truth'' was assumed and all the philosophical thinking was ordered afterwards. In other words, and as an outstanding example, nobody questioned the existence of God in the Middle Ages, with {\it Meditationes de Prima Philosophia} a relationship between the Subject (man) and Object (God, world, phenomena) is established. It is the Subject who validates the inquire on the Object ({\it Discours de la M\'ethode}).

\subsection{The Empiricism of Locke, Berkeley and Hume}

The work of several important philosophers during the Scientific Revolution of the 17th century is at the center of a division that was already present, but which was eventually identified and defended by John Locke and George Berkeley, and later intensely criticized by David Hume.

The philosophical movement opposed to French Rationalism is the Empiricism, born in England, became famous with the work of John Locke (1632 - 1704) who supported the theory of {\it tabula rasa}, or, that at birth, the human mind is like a blank canvas later progressively impressed through experiences arising from the senses. For Locke, innate ideas do not exist and the entire Theory of Knowledge supported by a rationalistic Metaphysics is criticized at its base.

Philosophers George Berkeley (1685 – 1753) and David Hume (1711 – 1776) would later reduce the reality of things to their perception, denying them as autonomous beings. For these latter two radical empiricists, it would be necessary that the objects of experiment and observation exist outside the human mind, or independently of it, but this thought is not clear, on the contrary.

Philosopher George Berkeley (1685 – 1753) would later reduce the reality of things to their perception and Idea, denying them as autonomous beings. For Berkeley, matter does not exist, being only pure appearance. The things we perceive from the outside world are just ideas, pure contents of the mind. Berkeley opposes (taking Locke's Empiricism to the extreme) the opposite idea: objects only exist as ideas, after being observed, that is, observation will grant them reality as ideas. For Berkeley, science is more descriptive than explanatory, which is why he cannot be considered a dogmatist.

Finally David Hume radicalizes Berkeley's concepts even more and comes to question the validity of the Principle of Causality, which is not logically sustained. Hume assimilated the human mind to a “theatre” where different actors and situations appear successively, but without us knowing who they are or where they come from. Hume remains skeptical about the reality of physical objects, in the saga of Locke and Berkeley with which he forms a kind of ``Trinity'' that offers an opposite view to the more intuitive Realism. We shall see how this ends up affecting the attitude of the German Idealists and modern scientists.

\subsection{The German Idealism}

Inspired by Hume, the philosopher Immanuel Kant (1724 – 1804) states that, to be certain of things that are external to us, a criterion is necessary, namely: the awareness of our individual existence (heavily put into question by Hume). It is from our own conscience that we can discern what is external and internal to us, although this point is not far from Berkeley. These criteria are a necessary tool  to move forward in this article. Kant admits the existence of objects in the physical world external to the observer, but maintains that we will never be able to know their essence (the {\it noumena}, ``the thing in itself''), because there are some cognitive limit in our minds, that hinder the analysis of phenomena. Kant also sustains that the so-called synthetic {\it a priori} judgments do not depend on external objects,
and that our mind shapes the phenomenological world into time, space and causality. Only through them we can model the natural world. It must be added that for Kant Mathematics is an exploratory tool character of the Natural Sciences without needing empirical content. Locke and his successors could not be more opposed to this form of Kantian Idealism.

The brief historical and ontological resumption made clear concepts such as the metaphysical world, the phenomenological world, rationalism, empiricism, realism and idealism.
These concepts will be seen to reappear, transposed to contemporary Physics, and in particular to Quantum Mechanics. In fact, these positions, dualistic as they are, underlie two other concepts pertinent to the nature of the activities of contemporary physicists, which are usually classified (in a very general and somewhat imprecise way) into Experimental and Theoretical.

\section{Realism and Idealism}

Could the empirical concept of Reality and the Realistic and also the Idealistic categories be applied to physicists? Or put another way, can we roughly divide them between the ``Experimental'' and the ``Theoretical''?

It should be noted that for this simple and dualistic classification, the relationship between the observer and the idea of the observed object is the basic, essential criterion. For the Realistic/Experimental Physicist, his relationship with the observed object is direct, for the Idealist/Theoretical Physicist, his relationship with the object is a rational construct.
A previous issue that emerges through the Philosophy of Science is whether it would be correct to divide physicists into these two categories, Realists and Idealists. If  labeled as such, realistic physicists are those who only identify as real objects that can be effectively verified through the senses, but they are usually dogmatic and static as the thought of Platonic Metaphysics, in evident paradox. With the exploration of the microworld, this question stays in the background, since quantum objects cannot be observed through the senses.
On the other hand, Idealist physicists, vaguely identified here as ``theoretical physicists'', would be those who are not satisfied with what has already been scientifically demonstrated. By presenting new hypotheses about reality, they would open new ways to be scientifically confirmed or refuted.

Evidently this division is neither rigid nor immutable, but would theoretical-idealist physicists be
more ``realistic'' than observational-realist physicists? Based on the history of knowledge, this question can and should be reformulated.

A long way has been traversed since the Aristotelian heavens, divided between rational, universal, immutable, timeless, perfect and real lunar and supralunar domains, until the day Galileo Galilei pointed his telescope and discovered the "mountains" of the Moon, the satellites of Jupiter and other evidence that provoked a scientific and epistemological revolution: with the help of instruments the macrophysical and microphysical world were not as they appeared to be. With technological advancement, the world of phenomena reveals itself increasingly different from what we perceive through our natural senses, and much more so when our senses cease to have any direct participation in apprehending the world, such as in the study of the microphysical world.
This implies several questions, but a very general theory of knowledge question arises: we do not perceive the world as it really is.

It is necessary to point out that the word ``world'' is usually used with different extensions, and in the case as used here, it can be applied both in the micro and in the macro physical sense.
The history of knowledge reveals to us that Platonic-Aristotelian dogmatic scheme rests on a dualistic dialectic by essence: being and non-being; the real and the unreal; the true and the false; the perfect and the imperfect and so on. This type of dualistic dialectic has obvious usefulness and application for realist physicists. For them reasoning is perfect and has a solid internal criterion, namely, its opposite: the opposite of perfect is imperfect; the positive is the negative; the real is the unreal and so on, without any kind of in-between thinking. However, this type of reasoning is essential on several occasions (such as a laboratory test that must result in positive or negative) and also for the dynamics of this article.

\section{The Reality of the Physical World}

The discussion of the Reality of the physical world in the context of the evolution of classical philosophical thought goes through the previously exposed philosophical views, with all their variations and disputes. If applied to Physics, these philosophical categories can be reduced to Theoretical Physicists (Idealists) and Experimental Physicists (Realists) already in the beginning of the 20th century or even before. It is important to point out that the interpretation of measurable quantities, the role of experiment and cognition, the existence of a Metaphysical world and the rest of the subjects discussed does not run into any concrete obstacles, being rather a purely philosophical debate in Classical Physics, until the beginning of the 20th century.

However, classifications are radically different and more difficult with the development of Quantum Mechanics, applicable to the domain of microphysics. The attempt to extrapolate the classical notions to the very small created numerous conceptual problems and required a detailed “interpretation”, that is, the attribution of meaning to each element present in the theory. In fact, almost a century after its construction, Quantum Mechanics enjoys a very special status in Physics: although almost nobody doubts it, in the sense that it always produced correct predictions (even when they were anti-intuitive), its interpretation is subject to discussion. It should be noted that most physicists adhere to the so-called Copenhagen interpretation, although there are other well-developed possibilities as alternatives [7]. But this adhesion occurs more by omission than by a thoughtful methodological decision. The alternatives are not even mentioned in higher education courses, nor are the problems of interpreting Copenhagen presented. That is, the theory is an act of faith, pure dogmatism, for its own practitioners, remembering the difficult position that Leucippus himself found himself in when he supported the ontological principle of matter through the atomic theory 2,500 years ago.

Faced with the numerous failures of attempts to find a consensual basis for QM, a permanent attitude of in-depth analysis would be desirable. Not the existence and effectiveness of QM, which are well known, are questioned, but its peculiar character. The purest empiricism, pointing to distrust anything that is not an experimental result, and the Cartesianism that maintains that the fact that something is not observable does not prevent its knowledge, seem to collide irreversibly in the microphysical world. In fact, we do not know how QM works, a process that remains in the theoretical or ideal world (for example, the so-called collapse of the wave function in the act of measurement is not contained in the theory, but is considered the fundamental factor of the results ). We only know that QM works, and this is presented to us as a sufficient factor for its existence and reality. We could say that at this moment we are between Leucippus' proposal and the dogmatic character of Platonic-Aristotelian Metaphysics. Needless to say, this is enough for a good part of physicists, who are not concerned with these subtler questions.

\section{Quantum Mechanics and Reality}

As any other physical theory, Quantum Mechanics consists of a handful of concepts, a mathematical formalism, and an {\it interpretation} to make sense of the results. Up to the advent of QM, all the interpretation issues were solved more or less quickly in the Classical Physics domain. This is why sometimes we tend to think that the interpretation need is a unique feature of QM, although it is not.

The need to give an ``interpretation'' to Quantum Mechanics, and the development of many different versions, is an example of the difficulties inherent in this theory. As stated, in Classical Physics there were also controversies and in-depth discussions regarding the meaning of introduced physical quantities (e.g. entropy, electromagnetic potentials, etc.) but they were finally resolved satisfactorily. Quantum Mechanics, on the other hand, will already complete a century without it being clear to the entire scientific community how it should be interpreted or understood. In fact, there are several alternative "interpretations", some of them that try to recover realism, others that justify the {\it sui generis} nature and treatment of quantum objects, and some that are even more particular. In general, we can say that the basic elements of a quantum theory are several and need to be distinguished and considered separately for a better understanding.

The first feature is that the subject-object separation must be reconsidered. In Classical Physics, the subject (S) is the observer who experiences the outside world, which are the objects (O) of reality (R) studied through the observation of phenomena and their analysis, the latter central in Quantum Mechanics and all other physical theories. In Western Science, in the classical domain, this separation is so evident that it is neither discussed nor even mentioned. However, in several other systems of thought (Oriental philosophy, native Americans), the separation between the subject S and the external reality R is impossible: man cannot put himself outside Nature. Indeed, some interpretations of Quantum Mechanics identify this separation as the source of the fundamental discrepancy between experiments and {\it a priori} expectations.

The interface that allows knowing quantum phenomena (Epistemology) is often pointed out as responsible for existing problems in understanding the theory. In general any physical theory needs at least three essential elements to deal with the object of study (system). These are:

(1) A Logic based on the Aristotelian principles of identity, on Aristotelian Logic (Boolean) and on the separation between subject (S) and object (O). As the subject S is implicitly supposed to be separable from the system R, there is the implicit assumption that the underlying logic is Boolean (Aristotelian). Von Neumann [13] was one of those who insisted on the possible non-human logic of Quantum Mechanics, analogous to the case of non-Euclidean geometry. For example, in the class of statements of the type ``if p, then p or q'' the quantum version ``if the Schr{\"o}dinger equation is valid, then the system evolves according to it and the result of a measurement will be one of the eigenvalues is constantly formulated without practitioners noticing its inconsistency with ordinary logic.

(2) A consistent Algebra to manipulate and relate the basic objects (the wavefunction $|\Psi {\bigr>}$, etc.) and obtain predictions and quantitative results. This is different from the adopted Logic, and fundamental for the results to be compared with the experiments (for example, the inner product in Hilbert space allows calculating the probabilities of measuring each eigenvalue in a single measure).

(3) Finally, a Language (formal or informal), which brings as a corollary a semantic-semiology, not always properly examined. In addition to Heisenberg's remarkable remark (in Petersen [1]) ``our words do not fit'', this point in the broader context (in the sense of the adequacy of human languages to formulate scientific claims) has been explored by Wittgenstein [8,9] and others, and remains a subject of great importance in the Philosophy of Science, with a dramatic bias in the case of interpretations of Quantum Mechanics.

We want to stress that, accepting the risk of returning to Idealism, the very idea of the existence of objects to be studied in Reality R is not guaranteed in Quantum Mechanics. In the later version of the Copenhagen interpretation, Bohr even stated that Quantum Mechanics is not about reality, but about what can be said about phenomena [1], suggesting that the former is an epistemological theory. In several versions of Quantum Mechanics it is possible to find a lot of philosophical Idealism, that is, the idea that the world is modeled by our minds, or at least, that the observer's mind has a lot to do with that (although there is no explicit mention to mind in the Copenhagen interpretation). Thus, the empirical content of the theory adopts a comprehensive dimension, and cannot be ignored. The denial of a quantum Ontology, or even its dismissal, is diametrically opposed to the Realist attitude of the physicist in everyday life.

To better visualize the elements involved in the interpretations of Quantum Mechanics, we have suggested the construction of conceptual diagrams that illuminate the role of each one. The most relevant is the diagram that exemplifies the so-called Copenhagen interpretation, illustrated in Fig. 1.

\begin{figure}[h]
	\centerline{\includegraphics[width=9truecm]{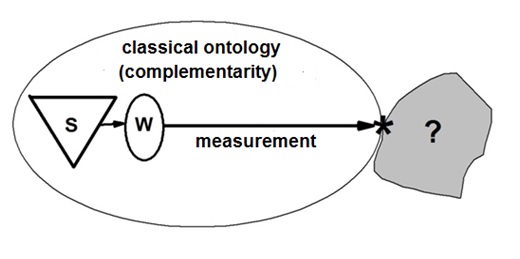}}
	\caption{A diagram representing the Copenhagen interpretation of quantum mechanics [2]. Subject S (triangle) measures a system and obtains a value represented by an asterisk. The relationship of the S with the measured phenomenon passes through the W filter, which contains the logic, the algebra and a language that expresses the relationship. Quantum reality itself is not specified, and falls into the category of the "thing in itself" that I. Kant discussed (grey zone with “?” sign). The Copenhagen interpretation maintains that the property measured with the asterisk is defined only at the moment of measurement (reality is created by the act of observation), and denies and {\it a priori} existence, in any trivial sense, to reality within the grey zone.}
\end{figure}

As an interesting episode that illustrates our discussion, years ago, one of the authors gave a seminar at IF-USP dealing with interpretations of QM. To his surprise, in the discussions, it became clear that among professional physicists nobody thought that QM is prescriptive and defines {\it a priori} what may or may not be the result of a measurement (called eigenvalue). That is, a dogmatic stance was accepted without much scrutiny, possibly because it works and inconsistent grounds are thus ignored. The Philosopher-Scientist (not to be confused with just a Philosopher of Sciences) Mario Bunge (1919 - 2020)  analyzed this and other aspects of the QM and clarified the difficulties arising from ``interpretation''. In this case, the Copenhagen orthodoxy that defines the results of a measure as an eigenvalue of a Hermitian operator, without the possibility of any other result, is blindly accepted [10].

As a final remark, we note that it is implicit in all scientist's discussions that physical Objects and Reality are the same thing, Kant's {\it noumena}. However, this view must be questioned by physicists when we take into account our cognitive limitations, i.e., space, time and causality relationships or even additional factors outside the Kantian thought.

\section{Quanta, locality and Bell's Theorem}

A vast literature has been devoted to the issue of the construction of Quantum Mechanics and its progressive departure from classical notions along the process. The new theory was not accepted at all by Einstein and other eminent physicists, including Schr{\"o}dinger and de Broglie. Their main objection was that a probabilistic outcome is not what Physics should offer, but rather a description of the Reality, i.e. quantum objects. They accepted, however, QM success, but insisted that the theory must be incomplete. Actually, they expected it to be superseded by a better theory, a more conventional one, for which QM emerged as some statistical smeared-out version.  It is fair to say that they hope to restore the situation in which quanta could be described as real entities with defined properties. Instead, the late versions of QM drifted towards an even more radical mood: Bohr stated at some point that QM was not about Reality, but rather about what can be said about Reality. In addition, the latter Reality was later plainly denied, in the sense that the QM formalism does not need any ``deep quantum Reality'', denying any ontological character to it and reducing the task to the study and prediction of phenomena. This attitude sharply contrasts with the statements attributed to Meyers, Einstein and others presented here and in next Section.

One of the most well-known attempts to advocate a realistic character to the physical objects and prove that Quantum Mechanics was incomplete is the work by Einstein, Podolsky and Rosen [17], in which they argue that it would be possible to bring in what they call ``elements of reality'' to achieve a full description of Nature. In fact, another important ingredient of the argument was the {\it locality} of the Reality. This kind of extended view is generally considered within the class of local hidden-variables [3], which should be introduced to bring QM to the realm of what Einstein and others found an acceptable description. In fact, the main objection from Einstein was the non-locality needed to make sense of quantum states, as needed by QM. D. Bohm developed a more precise definition of the system originally considered by EPR, using discrete spin states of positron-electron pairs. The main point of the whole argument is that pairs of quantum particles sent to, say, opposite directions are a superposition of spin states. The observers at locations A and B have spin detectors, and their measurements of the spin are made within a time interval $\Delta t$ which makes impossible for the first measured particle to communicate with the second with a velocity smaller than light $c$ (see Fig.2). Note that the Kantian categories of time, space and causality are
being used in full.

\begin{figure}[h]
	\centerline{\includegraphics[width=9truecm]{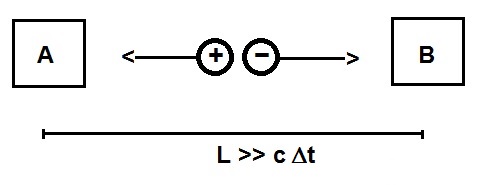}}
	\caption{Experimental setup using electron-positron pairs (``+'' and ``-'' at the center), later measured by two devices A and B widely separated, to guarantee that the information cannot causally reach the other once one of the spins is measured. The other one correlates perfectly, somewhat ``knew'' its brother was being measured far away from it. However, this is {\it not} a faster-than-light phenomenon}
\end{figure}

The problem is then ``what do the pairs of measurements show?''. Surprisingly (for Einstein, Podolsky and Rosen [17] at least) it is verified that the particles ``know'' what happens to their twins. If one direction of spin is measured for one, the other displays the opposite result, in spite that they are not causally connected because of the setup. It is said that particles are {\it entangled} (the phases of their wavefunction are correlated), but this entanglement does not depend on their relative distance and does not go away. In a sense, once the electron and positron formed a single system, they remain as such, independently of their distance. Sometimes it is stated that ``the world is an indivisible whole'' meaning the above entanglement. Another nickname for this found in the literature is ``spooky action-at-a-distance'' [5], meaning the type of non-locality behind the entanglement phenomenon.

Strange and unacceptable as this {\it gedankenexperiment} is (a favorite of Einstenian thought), actual experiments measuring entanglement were devised and performed, starting with the report of Aspect, Dalibard and Roger [16] and using several physical systems to test the predictions of QM. In fact, an important general development by [18] came up, establishing an upper bound to some directly measurable quantities within a local Reality approach as advocated by EPR. These bounds became known as {\it Bell's inequalities}. In several experiments using electron spins [19], as described above, but also trapped atoms [25], polarized photons [27], and Bose-Einstein ultracold helium atoms [26] Bell's inequalities were found to be violated at several standard deviations level, indicating that a local realistic theory of the type envisioned by EPR is not tenable.

What does this mean? Actually, a very serious fact: either a description of the microphysics could be constructed, but it must be {\it non-local}, or if local physics is kept, then the physical objects are not what we think they are. The rejection of non-local physics is so strong that some proposals have claimed that the world is {\it not} made of real objects, and even macroscopic objects are not there when nobody looks at them (vindicating the views of Berkeley [6] and the Idealism).

Of course, all kind of loopholes in this reasoning have been scrutinized, but no real refutation to this cumbersome situation, created by the violation of Bell's inequalities confirmed in many experiments, gained general acceptance. Entanglement is a fact not easily accommodated by a classical realistic attitude, but produces the results which seem as ``clean'' as they can be [19, 25], hence the importance of Bell's work and experimental confirmations of QM predictions.

\section{Why do Classical and Quantum Physics come into ontological conflict?}

We thus see the peculiarity of QM, which is accepted despite its dogmatic presentation, without much scrutiny, for the simple fact that it works, ignoring its inconsistent foundations. But this is not the end of the world: it takes time for the imagination to flow, to change, but this done, new theories will be presented, new antitheses and refutations will also be presented, until we arrive at new theories, new certainties, new realities, which also will become obsolete over time. Kronos is also relentless with his children's curiosity.

We are certain today that the separation of the physical from the philosophical disciplines was a gradual process that reached its peak in the 19th century. ``Natural philosophers'' began to be called ``biologists'', ``chemists'', ``physicists'', etc. to the extent of their specialization and refinement of the disciplinary matrix (the latter according to the definition of Kuhn [11]). A century after this bifurcation, we see that Science practitioners are no longer, in general, linked to Philosophy, nor do they feel obliged to respect its procedures, or to analyze their own work from its point of view, and its scope does not include perspectives philosophical in general.

Thus, the statement that every physicist is also a metaphysicist may even terrify many of them, but it was seriously expressed by A. Einstein, himself a philosopher-scientist as the title of the autobiographical book edited by Schlipp [28] says. Einstein observed that physicists need, even in an elementary and intuitive way, to adopt a metaphysics for the objects they study.

Einstein was the most famous opponent of the foundations of Quantum Mechanics, which maintains (in its orthodox Copenhagen interpretation) that there is no underlying ``quantum reality''. The theory predictions are thus operationalist, a kind of mere calculation algorithm, but they do not refer to Reality. According to this interpretation, the act of measuring a quantum system creates Reality, in the sense of forcing the system to define itself among several possibilities with probabilities expressed by the square of the wave function $|\Psi \bigr>$. Followers of this interpretation claim that this is all that can be obtained about the system, in stark contrast to classical physics. The idealist bias of the theory is quite clear, and in fact in other developed interpretations, this character is even more explicit and striking: for example, in the Wigner-von Neumann-Stapp interpretation it is stated that consciousness creates Reality [7, 14], in other words, fully aligns with Berkeley's thinking. It is possible to see why Einstein criticized Quantum Mechanics, which wants to get rid of the Metaphysics issue but falls into an unacceptable Idealism for Einstein and most of the physicists.

There are recent works [15] about Meyerson's epistemology and its influence on Einstein's so-called rationalist "turn" around 1930, and which serves to resume the question of Metaphysics (almost always involuntary) of professional physicists. In his work {\it Identit\'e et R\'ealit\'e} [20], Meyerson makes two major statements about the character of Science, as scientists themselves understand it. The first is that Science is explanatory, that is, that it is not reduced to describing and predicting phenomena, but intends to establish cause and effect relationships and reduce the observed/measured to general ideas. For this purpose, scientists reduce “common sense” in favor of an abstract view full of postulated objects (atoms, magnetic fields, fluxes, etc.) that are treated as real objects that exist independently of observation. This leads to the assertion that Science is ontological, that is, it prescribes the objects that exist. Hence the statement later repeated by Einstein himself that there is an inevitable fraction of Metaphysics in Physics. This general realist posture of physicists until the beginning of the 20th century was called ``intuitive'' by Meyerson [20], and suffered a serious blow with the emergence of Quantum Mechanics as we have just discussed.

This characteristic of considering abstract objects and mental constructs as ``real'' was taken up on several occasions, for example, in the Philosophy for Scientists course by Althusser and others, who calls it ``spontaneous'' [21]. Althusser discusses the process of building a philosophical posture in the scientist, but his motivation is historical-materialist, and he does not even mention Quantum Mechanics. An important work that approaches the scientific dynamics, its values and beliefs in a direct and extensive way is the book {\it Laboratory Life}, by B. Latour and S. Woolgar [22]. As an example, in this work the authors saw the neuropeptides studied in the laboratory of J. Salk in San Diego, USA as ``abstractions'' that embodied a series of signals from measuring devices, the only ``real'' evidence according to them, but which scientists they saw only as manifestations of those. The study led Latour and Woolgar to speak of Science as a ``social construction''. If that position taken seriously, Science has little or no relationship with Nature, being just a kind of game with its own rules. It is not even clear what is the reason why the manipulation of these concepts and abstract entities would, for example, with predictions of real systems, and also because Mathematics or even Logic would describe the phenomena of the World. Note the strange posture regarding this last point by none other than the Hungarian Nobel Prize winner E. Wigner, in his famous conference {\it The unreasonable effectiveness of Mathematics in the natural sciences} [23].

For Realistic/Experimental physicists, true Physics would be Realistic and the Philosophy of Science something totally dispensable. In fact, and in contrast to Einstein, some of the most eminent physicists of the 20th century have made no secret of their hostility towards philosophy. Just as an example, R.P. Feynman and P.A.M. Dirac seem not to have appreciated at all the philosophers' analysis of physical Reality, methodology, and other important aspects. In a quite crude and direct way (characteristic of the ``hard sciences'') they stated that Philosophy is at least superfluous, misleading and/or redundant. It should be said that the two, and in fact all the participants of the construction of the 20th century new Physics, had already chosen their Metaphysics (or lack of it): there are several writings where Dirac and Feynman manifest themselves in favor of Quantum Theory and consider the efforts of Einstein and others to restore Realism to the physical world to be useless. It should be noted that this position, even if correct, does not completely resolve the dualism discussed here.

\section{The epistemological dualism of Physics}

If until the 19th century Idealism could be considered an exotic doctrine, the construction of Quantum Mechanics in the first decades of the 20th century considerably complicates adherence to the more intuitive Realism, as we argued earlier. The revision of the Realist position was essential when Quantum Mechanics was built on a different basis. After a century, if on the one hand they adopt an operationalist philosophy (harshly criticized and discarded by Bunge [24]), physicists are horrified by ontological dilemmas explicitly expressed, but they accept without further a QM whose ontology is considered non-specified/non-existent for quantum objects (Copenhagen), that is, with idealistic connotations. When confronted with the macro-world, none of them (or almost) will admit, for example, that they disbelieve in the existence of objects while nobody observes them (which immediately takes us back to the philosophy of G. Berkeley once more), but that hovers in the later versions of Quantum Mechanics interpreted by the Copenhagen group, in the form ``the system has no definite properties if no one measures it''. That is, there is a flagrant philosophical duality, which is spreading and is never presented to new practitioners who arrive at Science (students) nor discussed among colleagues. It should be noted that this duality, even without being recognized and assimilated, is present in the same and only individual who practices hard science when he considers a quantum problem and a classical one. Without realizing it, he is forced to pass from an Idealist position to a Realist one, ignoring the inconsistencies of this process: the Ontology of objects does not admit an abrupt Idealist-Realist transition, one can pass from the quantum world to the classical limit, but not from a microphysical Idealism to a macrophysical Realism. This is the central point of this essay.

We can also state that, most of the time, it is even impossible to confront physicists to make them understand the situation, since they can adopt an attitude of disdain or even a radical skepticism, such as the one mentioned from Dirac and Feynman of Section 7. The idolatry of the hard sciences will prevent them from contradicting the founding fathers, and so it is even beneficial that they are not even aware of these last statements.

As a final summary of the problem, we have a dual and inconsistent position between quantum and classical phenomena that goes unnoticed. Either we find an interpretation (different from the Copenhagen orthodoxy in its late version) for Quantum Mechanics, which allows us to approach it and make it compatible with classical Realism, or we will be forced to claim at least partially some kind of Idealism for the classical world . For the authors, the first option is preferred, but we are far from having a consensus regarding the type of interpretation required, a delicate matter that far exceeds the scope of this essay [7, 14].

\bigskip
\noindent
{\it [I would] rather discover one cause than gain the kingdom of Persia.}

Democritus, quoted by K. Freeman (1948)

\bigskip
\noindent
{\it And yet, I still dream the dogmatic slumber of Immanuel Kant}

J.E. Horvath


\section*{Acknowledgments}

This work was performed under the auspices of a Research Fellowship granted by the {CNPq Agency}, Brazil
and {FAPESP Agency}, S\~ao Paulo State through the grant 2020/08518-2.

\section{References}

\noindent
1] Petersen, A. (1963), {\it Bull. Atomic Sci. }, {\it 19}, 8

2] Horvath, J.E. and Rosas Fernandes, R. (2023), {\it Conceptual diagrams in Quantum Mechanics}, submitted

3] Goldstein, S. (2021), {\it Bohmian Mechanics}{\it Stanford Encyclopedia of Philosophy} https://plato.stanford.edu/entries/qm-bohm/

4] Zilioli, U. (2020), {\it Atomism in Philosophy: A History from Antiquity to the Present }, Bloomsbury, UK

5] Mermin, D. (2007), {\it Quantum Computer Science: An Introduction}, Cambridge University Press, UK

6] Russell, B. (1989), {\it Wisdom of the West}, Crescent Books, USA

7] de la Pe\~na, L. (2010), {\it Introduction to Quantum Mechanics}, Ediciones Cient\'\i ficas Universitarias, M\'exico

8] Wittgenstein, L. (1954), {\it Tractatus Logico-Philosophicus }, Rutledge, NY, USA

9] Wittgenstein, L. (1953), {\it Philosophical Investigations }, McMillan Publishing, UK

10] Messiah, A. (2014), {\it Quantum Mechanics}, Dover Publications, UK

11] Kuhn, T. (1996), {\it The Structure of Scientific Revolutions }, The University of Chicago Press, Chicago, USA

12] Einstein,, A. (1998), {\it Albert Einstein, Philosopher-Scientist (translated and edited by A. Schlipp)}, Open Court, UK

13] von Neumann, J. (1996), {\it Mathematical Foundations of Quantum Mechanics}, Princeton University Press, USA

14] Johansson, L.-G. (2007), {\it Interpreting Quantum Mechanics},Taylor \& Francis, UK

15] Giovanelli, M. (2018), {\it Eur. Jour. Phil. Sci.},{\it 8}, 783

16] Aspect, A. and Dalibard, J. and Roger, J. (1982), {\it Phys. Rev. Lett.},{\it 49}, 1804

17] Einstein, A. and Podolsky, B. and Rosen, N. (1935), {\it Phys. Rev.},{\it 47}, 777

18] Bell, J.S. (1964), {\it Physics Physique Fizika},{\it 1}, 195

19] Hensen, B. et al. (2015), {\it Nature},{\it 526}, 682

20] Meyerson, E. (1908), {\it Identit\'e et r\'ealit\'e }, Alcan, Paris, France

21] Althusser, L. (2012), {\it Philosophy and the Spontaneous Philosophy of the Scientists: And Other Essays}, Verso Books, UK

22] Latour, B. and Woolgar, S.  (1986), {\it Laboratory Life}, Princeton University Press, NJ, USA

23] Wigner, E. (1960), {\it Comm. Pure Appl. Math. },{\it 13}, 1

24] Bunge, M. (2014), {\it Philosophy of Physics}, Springer, Berlin

25] Rosenfeld, W. et al. (2017), {\it Phys. Rev. Lett.},{\it 119}, 010402

26] Shin, D.K. et al. (2019), {\it Nature Comm. },{\it 10}, 4447

27] Jung, K. (2020), {\it Frontiers Phys. },{\it 8}, 170

28] Schlipp, A.. (1998), {\it Albert Einstein, Philosopher-Scientist (translated and edited by A. Schlipp) }, Open Court, UK

29] Freeman, K. (1948), {\it A Complete Translation of the Fragments in Diels, Fragmente der Vorsokratiker, 155}, Harvard University Press, USA

\end{document}